\documentclass[article]{ws-procs9x6}
\usepackage{amsmath,amsfonts,amssymb,epsfig,color}

\newcommand{\SU}[1]{\ensuremath{\mathrm{SU}( #1 )}}
\newcommand{\Un}[1]{\ensuremath{\mathrm{U}( #1 )}}
\newcommand{\SO}[1]{\ensuremath{\mathrm{SO}( #1 )}}

\newcommand{\Spn}[1]{\ensuremath{\mathrm{Sp}( #1 )}}
\newcommand{\SpR}[1]{\ensuremath{\mathrm{Sp}( #1,\mathbb{R} )}}
\newcommand{\su}[1]{\ensuremath{\mathfrak{su}( #1 )}}

\newcommand{\spn}[1]{\ensuremath{\mathfrak{sp}( #1 )}}
\newcommand{\spR}[1]{\ensuremath{\mathfrak{sp}( #1, \mathbb{R} )}}

\newcommand{\ket}[1]{\ensuremath{\left| #1 \right\rangle}}

\newcommand{\etal}{\emph{et al.}}

\newcommand{\IAS}{isobaric analog $0^+$ state}
\newcommand{\IASs}{isobaric analog $0^+$ states}
\newcommand{\fpg}{\ensuremath{1f_{5/2}2p_{1/2}2p_{3/2}1g_{9/2}} }

\newcommand{\flevel}{\ensuremath{1f_{7/2}} }

\newcommand{\upfp}{upper {\it fp}}

\begin{document}

\title{Fermion Systems with Fuzzy Symmetries\\
\scriptsize{(Leveraging the Known to Understand the Unknown)}
}
\author{\underline{J. P. Draayer}$^{1}$, K. D. Sviratcheva$^{1}$, \\
T. Dytrych$^{1}$, C. Bahri$^{1}$, K. Drumev$^{1}$, J. P. Vary$^{2}$
}
\address{$^{1}$Department of Physics and Astronomy, Louisiana State 
University, \\
Baton Rouge, Louisiana 70803, USA \\
$^{2}$Department of Physics and Astronomy, Iowa State University, \\
Ames, IA 50011, USA}

\maketitle

\abstracts{
Microscopic models, which embody the simplicity and significance of a
dynamical symmetry approach to nuclear structure, are reviewed. They
can reveal striking features of atomic nuclei when a symmetry dominates
and solutions in domains that may otherwise be unreachable.}

\noindent {\bf 1. Overview of algebraic fermion models.}
A theory that invokes group symmetries is driven by an expectation that
the wave functions of the quantum mechanical system under consideration can be
characterized by their invariance properties under the corresponding symmetry
transformations. But even when the symmetries are not exact, if one 
can find near
invariant operators, the associated symmetries can be used to help reduce the
dimensionality of a model space to a tractable size. Throughout the years,
group-theoretical approaches have identified fundamental symmetries in light
to heavy nuclei and achieved a reasonable reproduction of experimental data
(for a review of fermion models, see \cite{Draayer92}). In addition, they
provide theoretical predictions for nuclear systems including heavy unstable
nuclei not yet explored, and `exotic' nuclei, such as neutron-deficient or
$N\approx Z$ nuclei on the path of the nucleosynthesis $rp$-processes.

It is well-known that effective two-body interactions in nuclei are dominated
by pairing and quadrupole terms. The former gives rise to a pairing gap in
nuclear spectra, and the latter is responsible for enhanced electric quadrupole
transitions in collective rotational bands. Indeed, within the framework of the
harmonic oscillator shell-model, both limits have a clear algebraic 
structure in
the sense that the spectra exhibit a dynamical symmetry.  In the pairing limit
the symplectic \Spn{4} ($\sim\SO{5}$\cite{Helmers,HechtGinocchio}) 
group together
with its dual \Spn{2\Omega}, for $2\Omega$ shell degeneracy, use the seniority
quantum number\cite{Racah,Flowers} to classify the spectra.  On the other hand,
in the quadrupole limit the SU(3) (sub-) structure\cite{Elliott} governs a
shape-determined dynamics.

In light deformed nuclei, $A\lesssim 28$, the Elliot's \SU{3} 
model\cite{Elliott},
which incorporates the particle quadrupole and angular momentum operators,
proved successful for a microscopic description of collective modes. Indeed,
\SU{3}  is the exact symmetry group of the spherical oscillator, which is a
reasonable approximation for the average potential experienced by nucleons in
nuclei. Also, \SU{3} is the dynamical symmetry group of the deformed 
oscillator,
when, as is usually the case, the deformation is generated by quadrupole
interactions. In many cases, a single-irrep or few-irreps 
calculations suffice to
achieve good agreement with experimental rotational energy spectra and
electromagnetic transitions (e.g., see \cite{NaqviD90,VargasHD01}).

Limitations due to the fact that the \SU{3} model is applied within a shell and
in turn requires  effective charges for transition strengths are 
overcome in the
\SpR{3} symplectic shell model for light nuclei (for a review see 
\cite{Rowe85}).
It embeds the \SU{3} symmetry and in addition introduces important inter-shell
excitations, including high-$\hbar \omega$ correlations  and core excitations.
The symplectic shell model is a microscopic formulation of the Bohr-Mottelson
collective geometric model with a direct relation between a second- and a
third-order scalar products of the quadrupole operator and the ($\beta $,
$\gamma $)-shape variables. The \SpR{3} symplectic model provides a microscopic
description of monopole and quadrupole collective modes in deformed 
nuclei and a
reproduction of experimental rotational energy spectra and electromagnetic
transitions without effective charges (e.g., see
\cite{RosensteelR77,DraayerWR84}).

Furthermore, in the domain of light nuclei one can combine the \SpR{3}
symplectic shell model and the no-core shell model (NCSM)\cite{NavratilVB00} to
push forward the present frontiers in nuclear structure physics. The 
NCSM+\SpR{3}
allows us to use modern realistic interactions without any 
approximation (for the
interaction and the size of the model space) for low-$\hbar \omega $
configurations and hence to fully account for important short- and
intermediate-range correlations, while selecting only dominant 
high-$\hbar \omega
$ basis states responsible for multi-shell development of collective motion.

In the region of medium mass nuclei around the $N=Z$ line (currently 
explored by
radioactive beam experiments)  protons and neutrons occupy the same major shell
and hence their mutual interactions are expected to strongly influence the
structure and decay modes of such nuclei. In addition to 
like-particle ($pp$ and
$nn$) pairing correlations the close interplay of $pp$, $nn$ and proton-neutron
($pn$) pairs and the isospin-symmetry influence are microscopically 
described by
the \Spn{4} pairing model\cite{SGD04}.

For heavy nuclei ($A\gtrsim 100$), the discovery of the pseudo-spin
symmetry\cite{HechtA69,ArimaHS69} and its fundamental
nature\cite{Draayer91,BahriDM92,BlokhinBD95} establishes the pseudo-\SU{3}
model\cite{RatnaRajuDH73}. The pseudo-spin scheme is an excellent 
starting point
for a many-particle description of heavy nuclei, whether or not they are
deformed. As for the \SU{3} shell model, in many cases leading-irrep 
calculations
(e.g., see \cite{DraayerW83}) or mixed-irrep calculations (e.g., see
\cite{PopaHD00}) achieve good agreement with experimental data. The 
pseudo-\SU{3}
shell model provides a further understanding of the $M1$ transitions in nuclei
such as the even-even $^{160-164}$Dy and $^{156-160}$Gd isotopes, 
specifically it
reflects on the scissors and twist modes as well as the observed fragmentation,
that is, the break-up of the $M1$ strength among several levels 
closely clustered
around a few strong transition peaks in the 2-4 MeV energy
region\cite{BeuschelDRH98}.

In medium-mass and heavy nuclei, where the pseudo-spin scheme can be applied to
the normal parity orbitals and valence spaces are intruded by a unique parity
highest-$j$ orbit from the shell above, a major step towards understanding the
significance of the intruder level is achieved by a pseudo-\SU{3} plus intruder
level shell model\cite{WeeksHD81} that is currently under
development\cite{DrumevBGD04}.

Furthermore, the advantages of the symplectic \SpR{3} extension of the \SU{3}
model can be employed beyond the light nuclei domain towards a description of
heavy nuclei in the framework of the pseudo-\SpR{3} shell model (e.g., see
\cite{CastanosHDR91}). While early developments demonstrate the potential of
such a model for studying the structure of heavy nuclear systems, it has not
been fully explored.

In what follows, we present some recent results for three algebraic fermion
models where the symmetries are fuzzy -- meaning not exact -- but nevertheless
extremely useful in gaining a deeper understanding of the structure of real
nuclei.
\newline

\noindent {\bf 2. Symplectic \Spn{4} pairing model.}
An algebraic approach, with \Spn{4} the underpinning symmetry
and with only six parameters, can be used to provide a reasonable microscopic
description of pairing-governed $0^+$ states in a total of 306 even-even and odd-odd nuclei 
with mass $40\le A\le 100$ where protons and neutrons are filling the same major
shell\cite{SGD03,SGD04}. We employ the most general Hamiltonian with \Spn{4} dynamical symmetry,
\begin{eqnarray}
H_{\spn{4}} &=&
\textstyle{
-G\sum _{i=-1}^{1}\hat{A}^{\dagger }_{i}\hat{A}_{i}
-F \hat{A}^{\dagger }_{0}\hat{A}_{0}
-\frac{E}{2\Omega} (\hat{T}^2-\frac{3\hat{N}}{4 })
}
\nonumber \\
&-&
\textstyle{
D(\hat{T}_{0}^2-\frac{\hat{N}}{4})-C\frac{\hat{N}(\hat{N}-1)}{2}
-\epsilon \hat{N}
},
\label{clH}
\end{eqnarray}
where $G,F,E,D,C$ and $\epsilon >0$ are parameters (refer to Table
I in \cite{SGD04} for their estimates). In (\ref{clH}), $\hat{N}$ counts the total number of
particles, $\hat{T}^2$ is the isospin  operator and the
$\hat{A}^{\dagger }_{0,+1,-1}$ group generators, which build the 
basis states by
a consequent action on the vacuum state (a core like $^{40}$Ca or 
$^{56}$Ni), create, respectively, a proton-neutron ($pn$) pair, a proton-proton ($pp$)
pair or a neutron-neutron $(nn)$ pair of total angular momentum
$J^{\pi}=0^+$ and isospin $T=1$. The model Hamiltonian (\ref{clH}) 
represents an
effective microscopic interaction that conserves the $T_0$ third
projection of the isospin and includes proton-neutron and like-particle isovector ($T=1$) pairing
plus symmetry terms (the latter is related to a proton-neutron isoscalar ($T=0$) force). The model
interaction (\ref{clH}) is found to  correlate strongly with realistic interactions like
CD-Bonn+3 terms\cite{PopescuSVN05} in the
\flevel region\cite{SGD06} and reflects a large portion of the GXPF1 realistic
interaction\cite{HonmaOBM04} in the upper-$fp$ shell. In addition, 
the  $D$-term
in (\ref{clH}) introduces isospin symmetry  breaking and the $F$-term accounts
for a plausible, but weak, isospin mixing\cite{SGD04IM}. Both terms are
significant in non-analog
$\beta$-decays studies and also yield quantitative  results that are better
(e.g., by $85\%$ in $\flevel$) than the ones with $F=D=0$.

\begin{figure}[t]
\centerline{\\hbox{\epsfig{figure=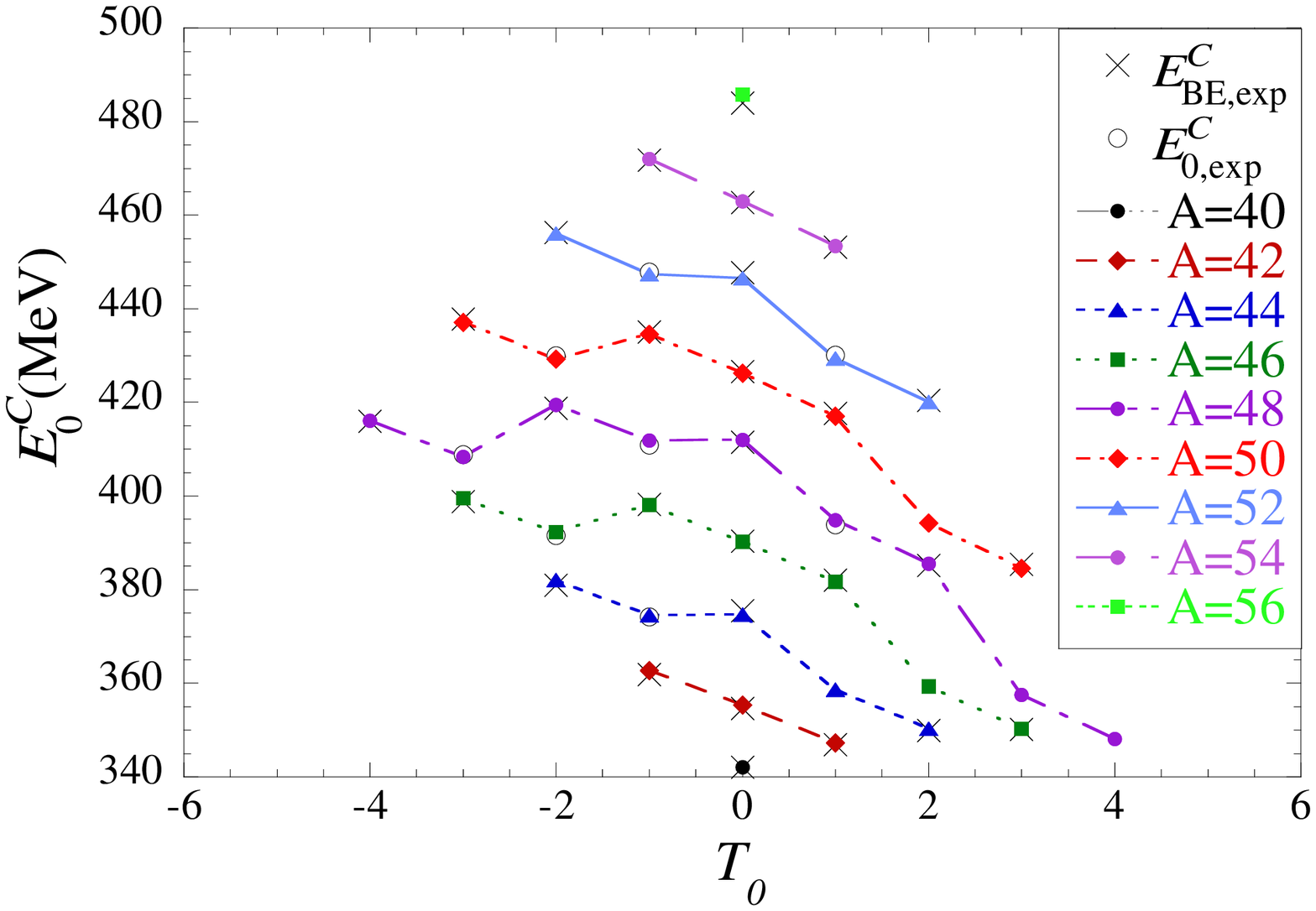,height=4.8cm,width=5.7cm}
\hspace{-12.4cm}\epsfig{figure=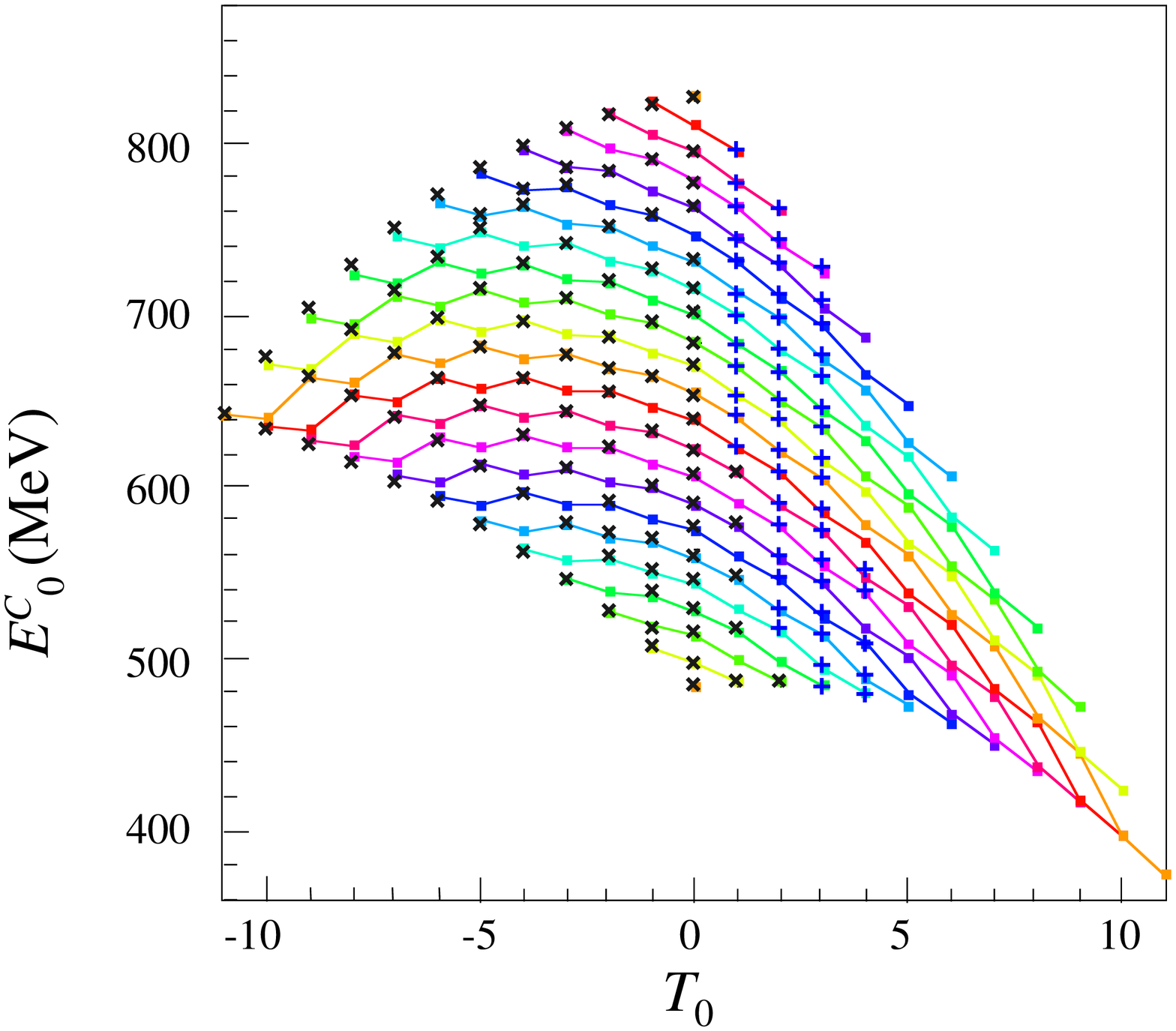,height=4.8cm,width=7cm}}}
\caption{Theoretical isobaric analog $0^{+}$ state energies in MeV of isobaric
sequences (lines)  (including the Coulomb energy) vs. the isospin projection
$T_0$.  Left: isobars with $A=56,58,\dots ,100$  in the \fpg shell ($^{56}$Ni
core), compared to experimental values (black `$\times $') and semi-empirical
estimates by P. M\"{o}ller \etal~ (blue `$+$'). Right: isobars in the \flevel
level compared to experimental binding energies ($\times$) or energies of the
lowest isobaric analog $0^{+}$ excited states ($\circ $).}
\label{CaE0expTh}
\end{figure}

Good agreement with experiment (small $\chi $-statistics) is observed
(Fig. \ref{CaE0expTh}) when theoretical eigenvalues of (\ref{clH}) are compared
to Coulomb-corrected\cite{RetamosaCaurier} experimental
energies\cite{AudiWapstraFirestone}. The theory predicts the lowest 
\IAS~  energy
with a deviation ($\chi /\Delta E_{0,exp}\times 100 [\% ]$) of $0.5\%$ for
\flevel and \fpg nuclei in the corresponding energy range considered, $\Delta
E_{0,exp}$.
The model estimates the binding energy of the proton-rich $^{48}$Ni nucleus
to be $348.1$ MeV, which is $0.07\%$ greater than the sophisticated
semi-empirical estimate\cite{MollerNK97} and only 4\% away from the 
experimental
value reported later-on\cite{Blank00}. The $^{68}$Se waiting-point 
nuclide along
the $rp$-path is estimated to be $574.3$MeV, only 2.7\% away from the 2004
precise mass measurement\cite{Clark04}. Likewise, for the odd-odd nuclei with
energy spectra not yet measured the theory predicts the energy of their lowest
isobaric analog
$0^{+}$ state:
$358.62$ MeV ($^{44}$V), $359.34$ MeV ($^{46}$Mn), $357.49$ MeV ($^{48}$Co),
$394.20$ MeV ($^{50}$Co) (Fig.
\ref{CaE0expTh}, right). The \Spn{4} model predicts the relevant $0^+$ state
energies for an additional 165 even-$A$ nuclei in the medium mass region (Fig.
\ref{CaE0expTh}, left). The binding energies for 25 of them are also calculated
in \cite{MollerNK97}. For these even-even nuclei, we predict binding 
energies that
on average are $0.05\%$ less than the semi-empirical
approximation\cite{MollerNK97}.

Furthermore, without any parameter variation, the theoretical energy spectra of
the \IASs~ are found to agree remarkably well with the experimental 
values where
data is available\cite{SGD04}. This agreement represents a valuable result
because the higher-lying $0^+$ states under consideration constitute an
experimental set independent of the data that enters the statistics 
to determine
the model parameters in (\ref{clH}).

In addition, we examine the detailed features of nuclei by discrete derivatives
of the energy function (\ref{clH}) filtering out the strong mean-field
influence\cite{SGD03stg}. This investigation reveals a remarkable 
reproduction of
the two-proton $S_{2p}$ and two-neutron $S_{2n}$ separation  energies, the
irregularities found around the $N=Z$ region, the like-particle and $pn$
isovector pairing gaps, and a prominent staggering behavior observed between
groups of even-even and odd-odd nuclides\cite{SGD03stg}.  The zero point of
$S_{2p}$ along an isotone sequence determines the two-proton-drip line, which
according to the $Sp(4)$ model for the \fpg  shell lies near the following
even-even nuclei\cite{SGD03stg}:
$^{60}$Ge$_{28}$,  $^{64}$Se$_{30}$,  $^{68}$Kr$_{32}$,  $^{72}$Sr$_{34}$,
  $^{76}$Zr$_{36}$,
$^{78}$Zr$_{38}$,  $^{82}$Mo$_{40}$,  $^{86}$Ru$_{42}$,  $^{90}$Pd$_{44}$,
  $^{94}$Cd$_{46}$,
beyond which the higher-$Z$ isotones are unstable with respect
to diproton emissions in close agreement with other
estimates\cite{Ormand97,MollerNK97,BrownCSV02} despite the lack of experimental
data. In addition, we find a small quadratic mean of the difference in
$S_{2p}$ between our model and the other theoretical predictions where data is
available, namely, $0.32$ MeV in comparison with \cite{Ormand97}, 
$0.78$ MeV with
\cite{MollerNK97} and 0.43 MeV with \cite{BrownCSV02}.

While the model describes only \IASs~of even-$A$
medium mass nuclei with protons and neutrons in the same shell, it reveals a
fundamental feature of the nuclear interaction, which
governs these states. Namely, the latter possesses a clear \Spn{4} dynamical
symmetry.
\newline

\noindent{\bf 3. Pseudo-SU(3) plus intruder level model.}
The role of intruder levels that penetrate down into lower-lying shells in
atomic nuclei has been the focus of many studies and debates. These levels are
found in heavy deformed nuclei where the strong spin-orbit interaction destroys
an underlying harmonic oscillator symmetry of the nuclear mean-field potential.

We carry out $m$-scheme shell-model calculations for the $^{58}$Cu
and $^{64}$Ge nuclei in the \fpg model space assuming
the occupancy of the $f_{7/2}$ orbital to be `frozen'. This choice was
motivated by the $f_{7/2}$ orbit's high occupation as reported
elsewhere\cite{Honma}. The Hamiltonian we used is a $G$-matrix with a
phenomenologically adjusted monopole part\cite{Retamosa}. A renormalized
version of this interaction in the $pf_{5/2}$ space has been introduced for
describing beta decays\cite{Van Isacker}.
\begin{figure}[thb]
\centerline{
\includegraphics[width=3.4in]{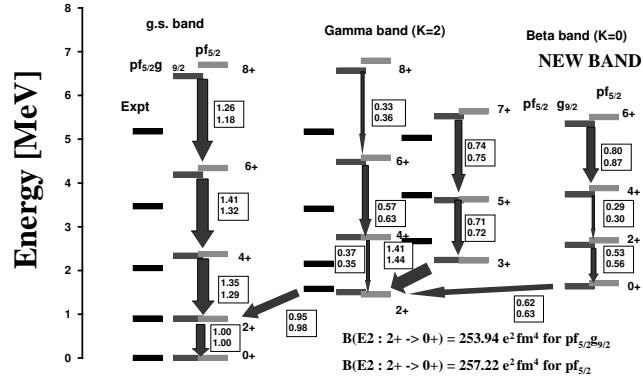}
}
\caption{Energy spectrum and B(E2) transition strengths for $^{64}$Ge. The
width of the arrows in the figure represent the relative B(E2) strengths,
normalized to unity for the ground band $2^+ \rightarrow 0^+$ transition.
Numbers in each box are for the \fpg and the restricted
\upfp~ model space, respectively.}
\label{su3intruderGe64}
\end{figure}

Results for the energy spectrum and B(E2) strengths of $^{64}$Ge are shown
in Fig. \ref{su3intruderGe64} for both model spaces. The  renormalized version
of the theory in the \upfp~ subspace not only reproduces the 
excitation energies
obtained in the larger \fpg space, but also gives very
similar values  for the B(E2) transition strengths. These results also
confirm those from a study using a schematic interaction\cite{Kaneko}.
Similar behavior was observed for $^{58}$Cu. Besides the ground state and
gamma bands for $^{64}$Ge a new (possibly beta) band is identified.

In addition, after rescaling occupancies of the states obtained in the \upfp~
subspace, a pattern that is very similar to that of the occupancies 
in the larger
\fpg space is obtained.

In short, novel shell-model calculations for $^{58}$Cu and $^{64}$Ge 
in the \fpg
model space using a realistic interaction are compared to those
generated by an appropriately renormalized counterpart of the interaction in
the truncated \upfp~ subspace. The results suggest that reliable
computations can be performed in a space that does not explicitly include the
intruder level as long as the interaction and the transition operators are
renormalized appropriately.
\newline

\noindent {\bf 4. No-core shell model plus \SpR{3} extension.}
The symplectic shell model is based on the noncompact symplectic 
\spR{3} algebra
with a subalgebraic structure that gives rise to rich underlying physics  for a
microscopic description of multiple collective  modes in nuclei. This follows
from the fact that the mass quadrupole and monopole moments operators, the
many-particle kinetic  energy, the angular and vibrational momenta are all
elements of the $\spR{3}\subset \su{3}$ algebraic structure\cite{Rowe85}.

The symplectic basis states, $|\Gamma_\sigma 
\Gamma_n\rho\Gamma_\omega \kappa (LS) J M_{J}\rangle$,
are constructed by acting with polynomials in the symplectic raising operator,
$A^{(2\,0)}$, on a set of basis states of a symplectic bandhead,
$\ket{\Gamma_{\sigma}}$. They are labeled according to the reduction chain
\begin{equation}
\begin{tabular}{ccccc}
$\SpR{3}$              & $\supset $     & $\Un{3}$              & 
$\supset$  & $\SO{3}$ \\
$\Gamma_\sigma\;\;\;\;$& $\Gamma_n\rho$ & $\;\;\;\Gamma_\omega$ & 
$\kappa$  & $L$
\end{tabular}
\nonumber
\end{equation}
where $\Gamma_\sigma\equiv N_\sigma \left(\lambda_{\sigma},
\mu_{\sigma}\right)$ labels \SpR{3} irreducible representations. The
$\Gamma_{n}\equiv n\left(\lambda_{n}, \mu_{n}\right)$  set of quantum numbers
gives the overall \SU{3} coupling of
$n/2$ raising operators acting on $\ket{\Gamma_{\sigma}}$.
$\Gamma_\omega\equiv N_\omega\left(\lambda_{\omega}, \mu_{\omega}\right)$
specifies the \SU{3} symmetry of a symplectic state and 
$N_\omega=N_\sigma+n$ is
the total number of oscillator quanta related to the eigenvalue,
$N_\omega \hbar\Omega$, of a center-of-mass motion free 
harmonic-oscillator (HO)
Hamiltonian. The basis states of a \SpR{3} irreducible  representation can be
expanded in HO ($m$-scheme) basis, which is the basis utilized by the no-core
shell-model (NCSM)\cite{NavratilVB00}.

In the case of $^{12}$C one can construct $13$ unique so-called
$0\hbar\Omega$-\SpR{3} irreducible representations. $0\hbar\Omega$ 
means that the
symplectic bandhead basis states, $\ket{\Gamma_{\sigma}}$, lie within
$0\hbar\Omega$ many-particle harmonic-oscillator space. For each of the
$0\hbar\Omega$-\SpR{3} irreducible representations we generate basis 
states up to
$N_{max}=6$ ($6\hbar\Omega$), which is the current  limit for NCSM 
calculations.
Typical the dimension of a symplectic representation is of order $10^{3}$,
comparing  to $10^{7}$ in the case of NCSM $m$-scheme basis space.
\begin{table}
\tbl{Probability distribution of the three most important \SpR{3} irreps and of
the NCSM wavefunctions.}
{\footnotesize
\begin{tabular}{c|rrrr|r}
\hline\hline
\multicolumn{6}{c}{~~~~~~~~~~~~~~J=0} \\
   & $0\hbar\Omega$ & $2\hbar\Omega$ & $4\hbar\Omega$ & $6\hbar\Omega$ 
& Total \\
\hline
	$(0\;4)S=0$ & $46.26$ & $11.39$ & $4.15$ & $1.11$ & $62.91$\\
	$(1\;2)S=1$ & $4.80$ & $1.87$ & $0.80$ &  $0.30$  & $7.77$\\
	$(1\;2)S=1$ & $4.72$ & $1.84$ & $0.79$ &  $0.29$  & $7.64$ \\
\hline
	\SpR{3} Total & $55.78$ & $15.10$ & $5.74$ &  $1.70$  & $78.32$ \\
  NCSM & $56.18$ & $22.40$ & $12.81$ &  $7.00$  & $98.38$ \\
\hline\hline
\multicolumn{6}{c}{~~~~~~~~~~~~~~J=2} \\
\hline
	$(0\;4)S=0$ & $46.80$ & $11.33$ & $3.99$ & $1.06$ & $63.18$\\
	$(1\;2)S=1$ & $4.84$ & $1.65$ & $0.69$ &  $0.25$  & $7.43$ \\
	$(1\;2)S=1$ & $4.69$ & $1.60$ & $0.67$ &  $0.24$  & $7.20$ \\
\hline
	\SpR{3} Total & $56.33$ & $14.58$ & $5.35$ &  $1.55$  & $77.81$ \\
NCSM & $56.63$ & $21.79$ & $12.73$ &  $7.28$ & $98.43$ \\
\hline\hline
\multicolumn{6}{c}{~~~~~~~~~~~~~~J=4} \\
\hline
	$(0\;4)S=0$ & $51.45$ & $11.23$ & $3.71$ & $0.94$  & $67.33$\\
	$(1\;2)S=1$ & $3.04$ & $0.89$ & $0.35$ &  $0.12$   & $4.40$ \\
	$(1\;2)S=1$ & $3.01$ & $0.88$ & $0.35$ &  $0.12$   & $4.36$ \\
\hline
	\SpR{3} Total & $57.50$ & $13.00$ & $4.41$ &  $1.18$    & $76.09$ \\
  NCSM & $57.64$ & $20.34$ & $12.59$ &  $7.66$     & $98.23$ \\
\hline\hline
\end{tabular}
\label{TABLE_15MeV}
}
\end{table}

The lowest-lying eigenstates of $^{12}$C are calculated by NCSM approach using
the Many Fermion Dynamics (MFD) code\cite{VaryZ94_MFD} with the effective
interaction derived from the realistic JISP16 {\it NN} potential for oscillator
strengths of $\hbar\Omega=15$ MeV. The large overlaps of the symplectic states
with the NCSM wave functions for  $0$, $2$, $4$ and $6\hbar\Omega$ subspaces of
the $m$-scheme basis (Tables~\ref{TABLE_15MeV}) reveal that around 80\% of the
latter are symmetric under \SpR{3} transformations. Apparently, for all three
wave functions, the highest contribution comes from the leading, most deformed,
$(0\,4)S=0$ \SpR{3} irreducible  representation. This contribution gets higher
towards $J=4^{+}_{1}$, where  mixing due to other, less deformed, 
configurations
decreases.

Clearly, the $0^{+}$, $2^{+}$ and $4^{+}$ states, which are 
constructed in terms
of the three \SpR{3} irreps with probability amplitudes defined by the overlaps
with the NCSM wavefunctions, can be used as a quite good approximation for a
microscopic description of the $0^{+}_{gs}$, $2^{+}_{1}$ and 
$4^{+}_{1}$ states in
$^{12}$C. Within this assumption, the $B(E2:2^+_1 \rightarrow 
0^+_{g.st.})$ value
turns out to be as much as 81\% of the NCSM estimate.

In short, the low-lying states in
$^{12}$C are quite well explained by only three \SpR{3} irreps of 1098
symplectic states, that is only 0.003\% of the NCSM space dimension, with a
dominance of the most deformed $(0~4)S=0$ collective configuration. 
Our findings,
as a `proof-of-principle', suggests that a NCSM+\SpR{3} structure 
could allow one
to extend no-core calculations to higher $\hbar \omega$ and heavier nuclei.

\vskip .3cm

In summary, models based on exact or just good (broken but dominant) 
symmetries in
fermion systems can play a significant role in our understanding 
low-lying nuclear
structure; specifically, as shown here, in the development of collective
rotational motion and the formation of correlated pairs in nuclei. They also
allow us to truncate a model space to typically only a fraction the size
encountered in models that do not exploit what we have dubbed here as fuzzy
symmetries.
\vskip .2cm

\noindent {\bf Acknowledgments}

This work was supported by the US National Science Foundation, Grant Numbers 0140300 \& 0500291 and
the Southeastern Universities Research Association (SURA).


\begin{thebibliography}{0}

\bibitem{Draayer92}  J. P. Draayer, {\it Fermion Models}, in {\it 
Algebraic Approaches to Nuclear
Structure}, ed. R. Casten, (Harwood Academic Publishers), Ch. 7, 423 (1992).

\bibitem{Helmers}  K. Helmers, {\it Nucl. Phys.} {\bf 23}, 594 (1961).

\bibitem{HechtGinocchio}  K. T. Hecht, {\it Nucl. Phys.} {\bf 63}, 177(1965);
{\it Phys. Rev.} {\bf 139},
B794 (1965); {\it Nucl. Phys.} {\bf A102}, 11 (1967);
J. N. Ginocchio, {\it Nucl. Phys.} {\bf 74}, 321 (1965).

\bibitem{Racah}  G. Racah, {\it Phys. Rev.} {\bf 62}, 438 (1942);
{\it Phys. Rev.} {\bf 63}, 367
(1943).

\bibitem{Flowers}  B. H. Flowers, {\it Proc. Roy. Soc.} (London) {\bf
A212}, 248 (1952).

\bibitem{Elliott} J. P. Elliott, {\it Proc. Roy. Soc.} (London) {\bf
A245}, 128 (1958); {\bf A245}, 562 (1958);
J. P. Elliott and M. Harvey, {\it Proc. Roy. Soc.} (London) {\bf 
A272}, 557 (1963).

\bibitem{NaqviD90} H.A. Naqvi and J.P. Draayer, {\it Nucl. Phys.} 
{\bf A516}, 351 (1990).

\bibitem{VargasHD01} C.E. Vargas, J.G.Hirsch, and J.P.Draayer, {\it 
Nucl. Phys.} {\bf A690},
409 (2001) .

\bibitem{Rowe85} D.J. Rowe, {\it Rep. Prog. Phys.} {\bf 48}, 1419 (1985).
\bibitem{RosensteelR77} G. Rosensteel and D.J. Rowe, {\it Phys. Rev. 
Lett.} {\bf 38}, 10 (1977);
{\it Ann. Phys. N Y} {\bf 126}, 343 (1980).

\bibitem{DraayerWR84} J.P. Draayer, K.J. Weeks, and G. Rosensteel, 
{\it Nucl. Phys.} {\bf A 413},
215 (1984).

\bibitem{NavratilVB00} P. Navr\'{a}til, J.P. Vary, and B.R. Barrett, 
{\it Phys. Rev. Lett.} $\mathbf{84}$, $5728$ ($2000$).


\bibitem{SGD04}  K. D. Sviratcheva, A. I. Georgieva, and J. P. Draayer,
{\it Phys. Rev.} {\bf C70}, 064302 (2004).

\bibitem{HechtA69} K. T. Hecht and A. Adler,  {\it Nucl. Phys.} {\bf 
A137}, 129 (1969).

\bibitem{ArimaHS69} A.  Arima A, M. Harvey, and M K. Shimizu, {\it 
Phys. Lett.} {\bf 30B}, 517
(1969).

\bibitem{Draayer91} J.P. Draayer, in {\it Proceedings of Symmetries 
in Physics}, Cocoyoc, Mexico,
June 3-7, 1991, ed. A. Frank and K.B. Wolf (Springer-Verlag, Berlin).

\bibitem{BahriDM92} C. Bahri, J. P. Draayer, and S.A. Moszkowski, 
{\it Phys. Rev. Lett.} {\bf 68},
2133 (1992).

\bibitem{BlokhinBD95} A.L. Blokhin, C. Bahri, and J.P. Draayer, {\it 
Phys. Rev. Lett.} {\bf 74},
4149 (1995).

\bibitem{RatnaRajuDH73} R.D. Ratna Raju, J.R Draayer, and K.T. Hecht, 
{\it Nucl. Phys.} {\bf A202},
433 (1973).

\bibitem{DraayerW83} J.R Draayer and K.J. Weeks, {\it Phys. Rev. 
Lett.} {\bf 51}, 1422 (1983).

\bibitem{PopaHD00} G. Popa, J. G. Hirsch, and J. P. Draayer, {\it 
Phys. Rev.} {\bf C62}, 064313
(2000).

\bibitem{BeuschelDRH98} T. Beuschel, J. P. Draayer, D. Rompf, and J. G. 
Hirsch, {\it Phys. Rev.} {\bf
C57}, 1233 (1998).

\bibitem{WeeksHD81} K.J. Weeks, C.S. Han, and J.P. Draayer, {\it 
Nucl. Phys.} {\bf A371}, 19 (1981).

\bibitem{DrumevBGD04} K. P. Drumev, C. Bahri, V. G. Gueorguiev and J. 
P. Draayer, in {\it
Proceedings of Nuclear Physics, Large and Small: International 
Conference on Microscopic Studies of
Collective Phenomena}, Morelos, Mexico, April 19-22, 2004, eds. R. 
Bijker, R.F. Casten, and A.
Frank, Melville, New York, {\it AIP Conference Proceeding} {\bf 726}, 
219  (2004).

\bibitem{CastanosHDR91} O. Casta\~{n}os, P.O. Hess, J.P. Draayer, and 
P. Rochford, {\it Nucl. Phys.}
{\bf A524}, 469 (1991).

\bibitem{SGD03}  K. D. Sviratcheva, A. I. Georgieva, and J. P.
Draayer, {\it J. Phys. G: Nucl. Part.
Phys.} {\bf 29}, 1281 (2003).

\bibitem{PopescuSVN05} S. Popescu, S. Stoica, J. P. Vary, and P.
Navratil, to be published.

\bibitem{SGD06}  K. D. Sviratcheva, J. P. Draayer, and J. P. Vary,
{\it Phys. Rev.} {\bf C73}, 034324 (2006).

\bibitem{HonmaOBM04} M. Honma, T. Otsuka, B. A. Brown, and T. Mizusaki, {\it
Phys. Rev.} {\bf C69}, 034335 (2004).

\bibitem{SGD04IM}  K. D. Sviratcheva, A. I. Georgieva, and J. P.
Draayer, {\it Phys. Rev.} {\bf C72}, 054302 (2005).

\bibitem{RetamosaCaurier}  J. Retamosa, E. Caurier, F. Nowacki and A.
Poves, {\it Phys. Rev.} {\bf
C55}, 1266 (1997).

\bibitem{AudiWapstraFirestone}  G. Audi and A. H. Wapstra, {\it Nucl. 
Phys.} {\bf
A595}, 409 (1995); R. B. Firestone and C. M. Baglin, {\it Table of Isotopes},
8th Edition (John Wiley \& Sons, 1998).

\bibitem{MollerNK97}  P. M\"{o}ller, J. R. Nix and K.-L. Kratz, {\it
LA-UR-94-3898} (1994); {\it At. Data Nucl. Data Tables} {\bf 66}, 131 (1997).

\bibitem{Blank00} B. Blank \etal, {\it Phys. Rev. Lett.} {\bf 84}, 1116 (2000).

\bibitem{Clark04} J.A. Clark \etal, {\it Phys. Rev. Lett.} {\bf 92}, 
192501 (2004).

\bibitem{SGD03stg}  K. D. Sviratcheva, A. I. Georgieva, and J. P. Draayer,
{\it Phys. Rev. } {\bf C69}, 024313 (2004).

\bibitem{Ormand97} E. Ormand, {\it  Phys. Rev. } {\bf C55}, 2407 (1997).

\bibitem{BrownCSV02} B. A. Brown, R. R. C. Clement, H. Schatz,
and A. Volya,
{\it  Phys. Rev.} {\bf C65}, 045802 (2002).

\bibitem{Honma} M. Honma, T. Mizusaki, and T. Otsuka, {\it Phys. Rev. Lett.}
{\bf 77},  3315 (1996).

\bibitem{Retamosa} E. Caurier, F. Novacki, A. Poves, and J. Retamosa,
{\it Phys. Rev. Lett.} {\bf 77}, 1954 (1996).

\bibitem{Van Isacker}  P. Van Isacker, O. Juillet, and F. Nowacki, {\it Phys. Rev.
Lett.}  {\bf 82}, 2060 (1999).

\bibitem{Kaneko} K. Kaneko, M. Hasegawa, and T. Mizusaki, {\it Phys. Rev.}
{\bf C66}, 051306(R) (2002).

\bibitem{VaryZ94_MFD} J. P. Vary and D. C. Zheng, {\it ``The 
Many-Fermion-Dynamics Shell-Model Code"}, Iowa State University, 1994
(unpublished).

\end{thebibliography}
\end{document}